\documentclass{article}
\usepackage{spconf,graphicx,hyperref}
\usepackage{amsmath,amssymb,amsfonts}
\usepackage{textcomp}
\usepackage{xcolor}
\usepackage{url}
\usepackage{multirow}
\usepackage[normalem]{ulem}
\usepackage{subcaption}
\usepackage{booktabs}
\usepackage{caption}
\usepackage{graphicx}
\usepackage{algorithm}
\usepackage{algpseudocode}
\usepackage{amsmath}
\usepackage{tabularx}
\usepackage{makecell}
\usepackage{float}
\usepackage{cuted}
\usepackage{caption}
\usepackage{stfloats}
\title{Towards Realistic Synthetic Data for Automatic Drum Transcription}
\author{{\em Pierfrancesco Melucci$^{1,2}$, Paolo Merialdo$^{2}$, Taketo Akama$^{3}$}\\[1.0em]
$^{1}$Sapienza University of Rome, Italy; \\
$^{2}$Roma Tre University, Italy; \\
$^{3}$Sony Computer Science Laboratories, Tokyo, Japan}

\begin{document}

\maketitle
\begin{abstract}

Deep learning models define the state-of-the-art in Automatic Drum Transcription (ADT), yet their performance is contingent upon large-scale, paired audio-MIDI datasets, which are scarce. Existing workarounds that use synthetic data often introduce a significant domain gap, as they typically rely on low-fidelity SoundFont libraries that lack acoustic diversity. While high-quality one-shot samples offer a better alternative, they are not available in a standardized, large-scale format suitable for training.
This paper introduces a new paradigm for ADT that circumvents the need for paired audio-MIDI training data. Our primary contribution is a semi-supervised method to automatically curate a large and diverse corpus of one-shot drum samples from unlabeled audio sources. We then use this corpus to synthesize a high-quality dataset from MIDI files alone, which we use to train a sequence-to-sequence transcription model. We evaluate our model on the ENST and MDB test sets, where it achieves new state-of-the-art results, significantly outperforming both fully supervised methods and previous synthetic-data approaches.
The code for reproducing our experiments is publicly available at \url{https://github.com/pier-maker92/ADT_STR}.

\end{abstract}
\section{Introduction}
\label{sec:intro}
Automatic Drum Transcription (ADT) is the task of generating a symbolic music transcription from an audio recording of a drum performance. The task can be specified for several scenarios, including drum-only recordings (DTD), drums with other percussion (DTP), and drums within full musical arrangements (DTM) \cite{Wu2018-yw}. Since the adoption of Deep Neural Networks for ADT, a significant challenge has been the scarcity of large-scale, paired audio-MIDI corpora required for supervised training. As MIDI-only datasets are far more abundant, synthetic data generation has been widely explored as a solution \cite{Vogl2018-wj, Mezza2024-ft, Cartwright2018-rj}.
However, these synthetic approaches introduce a critical domain gap and effectively reframe the task as an out-of-distribution generalization problem. Indeed, previous works have shown that synthetic data is most effective when used in combination with real-world data, e.g. when pretrained models on synthetic datasets are subsequently finetuned on real-world distributions. This observation indicates that the distribution shift introduced by synthetic data is too large to rely solely on synthetic data generation during training. We hypothesize that this issue stems from inherent limitations of SoundFont-based synthesis, which tends to produce low-quality audio from MIDI representations.
High-quality one-shot drum samples provide a more realistic and diverse sound source, and their use has been shown to be effective for training models for music transcription \cite{Sato2023-ry}, but in the context of ADT their practical use is hindered by a lack of standardization. Publicly available one-shot libraries suffer from two main issues: $(i)$ ambiguous or incorrect labels for similar instruments (e.g., ride vs. crash cymbals), and $(ii)$ absence of a consistent naming convention, which must be inferred through further analysis of the spectral content, as proposed in \cite{Cartwright2018-rj}.
In addition to the synthetic-to-real distribution gap, we highlight another limitation in the current state of the art of ADT systems from an architectural perspective. Sequence-to-sequence models have been shown to perform very well in Music Information Retrieval (MIR) tasks, such as music transcription \cite{Gardner2021-bp, Sato2023-ry, Chang2024-dw}. However, these works report drum transcription results primarily for ablation purposes, with the evaluation of drum signals lying outside the main scope of the study, which is instead focused on the broader task of multi-instrument music transcription. Consequently, a systematic evaluation of this type of architecture when trained specifically for automatic drum transcription (ADT) tasks is still lacking.
Motivated by this, we investigate an efficient synthetic data generation pipeline that aims to reduce the distributional gap between synthetic and real data by incorporating one-shot sample libraries, as proposed in \cite{Sato2023-ry, Cartwright2018-rj}. We further validate our approach by training a sequence-to-sequence transformer specifically tailored for the ADT task. We summarize the contributions of this study as follows:
\begin{itemize}
\item We propose a scalable semi-supervised pipeline for creating a large, diverse, and standardised corpus of one-shot drum samples, starting from unstructured, publicly available one-shot libraries.
\item We empirically validate our approach by training a model exclusively on synthetic data that outperforms previous methods on real drum signal distributions, as evaluated on the ENST and MDB datasets.
\end{itemize}
\renewcommand{\arraystretch}{0.8}
\begin{table*}[t]
\centering
\scriptsize
\caption{MIDI Key to GM-Percussion Mapping}
\label{tab:midi_mapping}
\resizebox{\textwidth}{!}{%
\begin{tabular}{cccc|cccc}
\toprule
\textbf{Key Source} & \textbf{Label Source} & \textbf{Instrument ID} & \textbf{Label Target} &
\textbf{Key Source} & \textbf{Label Source} & \textbf{Instrument ID} & \textbf{Label Target} \\
\midrule
35 & Acoustic Bass Drum & 0  & Acoustic Bass Drum  & 58 & Vibraslap         & 18 & Vibraslap \\
36 & Bass Drum 1        & 1  & Bass Drum 1         & 59 & Ride Cymbal 2     & 13 & Ride Cymbal \\
37 & Side Stick         & 2  & Side Stick          & 60 & Hi Bongo          & 19 & Congas \& Timbales \\
38 & Acoustic Snare     & 3  & Acoustic Snare      & 61 & Low Bongo         & 19 & Congas \& Timbales \\
39 & Hand Clap          & 4  & Hand Clap           & 62 & Mute Hi Conga     & 19 & Congas \& Timbales \\
40 & Electric Snare     & 5  & Electric Snare      & 63 & Open Hi Conga     & 19 & Congas \& Timbales \\
41 & Floor Tom          & 6  & Floor Tom           & 64 & Low Conga         & 19 & Congas \& Timbales \\
42 & Closed Hi Hat      & 7  & Closed Hi Hat       & 65 & High Timbale      & 19 & Congas \& Timbales \\
43 & High Floor Tom     & 6  & Floor Tom           & 66 & Low Timbale       & 19 & Congas \& Timbales \\
44 & Pedal Hi-Hat       & 8  & Pedal Hi-Hat        & 67 & High Agogo        & 17 & Cowbell \\
45 & Low Tom            & 6  & Floor Tom           & 68 & Low Agogo         & 17 & Cowbell \\
46 & Open Hi-Hat        & 9  & Open Hi-Hat         & 69 & Cabasa            & 20 & Shaker \\
47 & Low Mid Tom        & 10 & Mid Tom             & 70 & Maracas           & 20 & Shaker \\
48 & High Mid Tom       & 10 & Mid Tom             & 71 & Short Whistle     & 21 & Whistle \\
49 & Crash Cymbal       & 11 & Crash Cymbal        & 72 & Long Whistle      & 21 & Whistle \\
50 & High Tom           & 12 & High Tom            & 73 & Short Guiro       & 22 & Guiro \\
51 & Ride Cymbal        & 13 & Ride Cymbal         & 74 & Long Guiro        & 22 & Guiro \\
52 & Chinese Cymbal     & 14 & Chinese Cymbal      & 75 & Claves            & 23 & Claves \\
53 & Ride Bell          & 13 & Ride Cymbal         & 76 & Hi Wood Block     & 23 & Claves \\
54 & Tambourine         & 15 & Tambourine          & 77 & Low Wood Block    & 23 & Claves \\
55 & Splash Cymbal      & 16 & Splash Cymbal       & 78 & Mute Cuica        & 24 & Cuica \\
56 & Cowbell            & 17 & Cowbell             & 79 & Open Cuica        & 24 & Cuica \\
57 & Crash Cymbal 2     & 11 & Crash Cymbal        & 80 & Mute Triangle     & 25 & Triangle \\
   &                    &    &                     & 81 & Open Triangle     & 25 & Triangle \\

\bottomrule
\end{tabular}
}
\end{table*}
\renewcommand{\arraystretch}{1.0}
\section{Related works}
\label{sec:format}
The field of ADT has been systematically reviewed in \cite{Wu2018-yw}, which outlined the primary transcription scenarios (DTD, DTP, DTM) and identified the Convolutional Recurrent Neural Network (CRNN) followed by a peak-picking algorithm as the dominant architectural paradigm. While alternative sequence-to-sequence Transformer models have been proposed \cite{Gardner2021-bp} for music transcription, in the context of ADT, only a few works have explored different architectures other than CRNNs, limiting the analysis to employing self-attention in such convolutional systems \cite{Zehren2023-ti}.  In response to the lack of paired audio–MIDI datasets, a substantial body of prior work has adopted data-centric approaches to address the data scarcity. Rather than focusing on architectural designs tailored to the ADT task, these studies primarily investigate how large-scale data can be leveraged through synthetic data generation strategies or crowd-sourced collections. Cartwright and Bello demonstrated that combining a small real dataset with a larger synthetic one improves performance, expanding the transcription vocabulary~\cite{Cartwright2018-rj}. Following this direction, Vogl \emph{et al.} introduced the TMIDT dataset, a large synthetic corpus, and found that pre-training on synthetic data before fine-tuning on a mix of real and synthetic data yielded the best performance for up to 18 instruments~\cite{Vogl2018-wj}. In contrast to synthetic methods, ADTOF~\cite{Zehren2021-ht}, a large-scale, non-synthetic dataset built using crowd-sourced annotations from rhythm games, showed that it was able to produce models with strong generalization. Further analyses have aimed to understand and mitigate the synthetic-to-real domain gap, identifying rhythmic quantization and compositional simplicity as key failure points~\cite{Zehren2024-kx}.
\section{Methodology}
\label{sec:pagestyle}
To address the data scarcity problem in ADT, we train a sequence-to-sequence transformer model using a pipeline that relies exclusively on synthetically generated audio, following \cite{Gardner2021-bp, Chang2024-dw, Sato2023-ry}. Our methodology comprises three main components: $(i)$ the curation of a one-shot labeling convention derived from the standard MIDI Percussion Key Map, resulting in a vocabulary of 26 drum instrument classes; $(ii)$ an unsupervised approach to automatically assign unseen one-shot samples to this 26-class schema based on their similarity to the curated dataset; and $(iii)$ an on-the-fly synthesis engine that ensures high diversity and substantial variability in the generated synthetic audio used for training.
\subsection{Instrument Vocabulary}
\label{ssec:subhead}
Our approach begins with establishing a standardized and comprehensive instrument vocabulary. As reported in Table~\ref{tab:midi_mapping}, we derived our vocabulary from the standard MIDI Percussion Map, consolidating the original 47 percussion notes into 26 distinct instrument classes by merging acoustically equivalent instruments, such as those that differ only by slight intonation variations (e.g. Crash Cymbal and Crash Cymbal 2, Mute Cuica and Open Cuica). This mapping simplifies the classification task while retaining the essential percussive elements required for detailed transcription. 
\subsection{Semi-Supervised One-Shot Library Creation}
\label{ssec:subhead}
\begin{algorithm}[t]
\caption{Semi-Supervised One-Shot Drum Sample Classification}
\label{alg:semi_supervised_oneshot}
\begin{algorithmic}

\Require 
Unlabeled one-shot sample set $U$ \\
Manually labeled seed set $G = \bigcup_{i=1}^{26} G_i$ with class set $C = \{c_1, \ldots, c_{26}\}$ \\
Audio encoder $E_A$

\Ensure 
Labeled dataset $D = \{(x, \hat{c}, \operatorname{conf}(x))\}$

\State \textbf{Compute class centroids}
\For{each class $c_i \in C$}
    \State $V_i \gets \emptyset$
    \For{each sample $g \in G_i$}
        \State $V_i \gets V_i \cup \{ E_A(g) \}$
    \EndFor
    \State $\mathbf{v}_i \gets \frac{1}{|V_i|} \sum_{\mathbf{v} \in V_i} \mathbf{v}$
\EndFor

\Statex
\State Initialize labeled dataset $D \gets \emptyset$

\Statex
\State \textbf{Classify unlabeled samples}
\For{each sample $u \in U$}
    \State $\mathbf{e}_u \gets E_A(u)$
    \For{each class $c_i \in C$}
        \State $S(u, c_i) \gets 
        \frac{\mathbf{e}_u \cdot \mathbf{v}_i}
        {\lVert \mathbf{e}_u \rVert \, \lVert \mathbf{v}_i \rVert}$
    \EndFor
    \State $\hat{c} \gets \arg\max_{c_i \in C} S(u, c_i)$
    \State $\operatorname{conf}(u) \gets \max_{c_i \in C} S(u, c_i)$
    \State $D \gets D \cup \{(u, \hat{c}, \operatorname{conf}(u))\}$
\EndFor

\Statex
\Return $D$

\end{algorithmic}
\end{algorithm}

To train a robust model for ADT, a large and diverse dataset of one-shot drum samples is essential. We constructed such a library of 8,495 samples through a semi-supervised classification pipeline. The process began with an unlabeled corpus of one-shot sounds aggregated from 12 public sample packs\footnote{http://soundpacks.com}. To impose structure upon this data, we first manually annotated a small seed set of 1,421 samples, creating a "gold" reference across 26 distinct instrument classes, denoted as the set \(C = \{c_1, c_2, \ldots, c_{26}\}\).
The core of our method lies in creating a representative prototype for each instrument class against which all other samples could be compared. For each class \(c_i\), we define its gold set as \(G_i\). We then leverage the rich feature space of the Contrastive Language-Audio Pre-training CLAP model \cite{Elizalde2022-xz}, using its audio encoder \(E_{A}\) to extract an embedding for each sample in \(G_i\). The resulting embeddings for each class were averaged to compute a single class centroid, \(\mathbf{v}_i\), which serves as the canonical representative for that instrument in the embedding space. This is formally expressed as:
\[
\mathbf{v}_i = \frac{1}{|G_i|} \sum_{g \in G_i} E_{A}(g)
\]
With these 26 class centroids established, we proceeded to classify the remaining 7,074 unlabeled samples. For each unlabeled sample \(u\), we first computed its CLAP embedding \(E_{A}(u)\). We then quantified its relationship to each instrument class by calculating a similarity score, \(S(u, c_i)\), between the sample's embedding and each class centroid \(\mathbf{v}_i\). This score is defined by the cosine similarity, which measures the cosine of the angle between the two vectors in the embedding space:
\[
S(u, c_i) = \frac{E_{A}(u) \cdot \mathbf{v}_i}{\|E_{A}(u)\| \, \|\mathbf{v}_i\|}
\]
This step produced, for each unlabeled sample, a vector of 26 scores, representing its semantic proximity to every defined instrument class. The final classification was determined by assigning the sample to the class for which it achieved the highest score. The class assignment, \(c^*\), is therefore given by:
\[
c^* = \arg\max_{c_i \in C} S(u, c_i)
\]
Crucially, the maximum similarity score itself provides a valuable metric for the quality of the classification. We retain this value as a confidence score, \(\operatorname{conf}(u)\), associated with each newly labeled sample. This confidence score is defined as:
\[
\operatorname{conf}(u) = \max_{c_i \in C} S(u, c_i)
\]
This score-driven, semi-supervised procedure allowed us to rapidly and automatically expand our initial seed set into a comprehensive library, with each sample possessing not only a class label but also a quantifiable confidence metric. This enables fine-grained control over data quality and diversity during subsequent model training phases.
We report the pseudocode for the procedure in Algorithm~\ref{alg:semi_supervised_oneshot}
\subsection{Tokenization and model architecture}
\label{ssec:Tokenization and Model}
\subsubsection{Tokenization}
\label{sssec: Tokenization}
We adopt an encoder-decoder Transformer \cite{vaswani2017attention} architecture to model the sequence-to-sequence mapping from audio features to symbolic MIDI tokens, following \cite{Gardner2021-bp, Sato2023-ry, Chang2024-dw}. 
To represent MIDI events as discrete tokens, we employ an adapted version of the standard MIDI tokenization proposed in previous works \cite{Amagasu2024-no, Gardner2021-bp, Chang2024-dw, Sato2023-ry}. Our MIDI token vocabulary is structured to capture three essential attributes of a musical event:
\begin{itemize}
\item \textbf{Tempo:} Represented by 245 distinct values, providing a high-resolution temporal quantization of 10 ms (1/100 seconds).
\item \textbf{Instrument Category:} Denoted by an ID corresponding to one of the 26 instrument classes defined in our vocabulary.
\item \textbf{Velocity (Optional):} Represented by 127 standard MIDI velocity levels, capturing the dynamic intensity of the event.
\end{itemize}
\subsubsection{Model architecture}
\label{sssec: Model architecture}
The encoder is fed with the mel-spectrogram of a 2.56-second audio window. This continuous input is processed by the encoder to extract a sequence of high-level feature embeddings. The decoder is then trained to autoregressively generate the corresponding sequence of MIDI tokens, effectively reconstructing the symbolic representation of the drum performance from the audio input. Formally, given an input audio segment $x$, the encoder maps the corresponding mel-spectrogram to a sequence of latent representations
\begin{equation}
\mathbf{H} = \mathrm{Enc}(x) = \{\mathbf{h}_1, \mathbf{h}_2, \ldots, \mathbf{h}_T\},
\end{equation}
where $\mathbf{h}_t \in \mathbb{R}^d$ denotes the encoder embedding at time step $t$. The decoder models the symbolic transcription as an autoregressive sequence of MIDI tokens
$\mathbf{y} = (y_1, y_2, \ldots, y_N)$, where each token $y_n$ belongs to the predefined vocabulary
$\mathcal{V}$ comprising tempo, instrument category, and (optionally) velocity tokens.
The conditional probability of the output sequence given the audio input is factorized as
\begin{equation}
p(\mathbf{y} \mid x) = \prod_{n=1}^{N} p(y_n \mid y_{<n}, \mathbf{H}),
\end{equation}
where $y_{<n} = (y_1, \ldots, y_{n-1})$ denotes the previously generated tokens.
Each conditional distribution is parameterized by the Transformer decoder via a softmax layer:
\begin{equation}
p(y_n \mid y_{<n}, \mathbf{H}) = \mathrm{softmax}\big( f_{\theta}(y_{<n}, \mathbf{H}) \big),
\end{equation}
with $f_{\theta}(\cdot)$ denoting the decoder network with parameters $\theta$. The model is trained using teacher forcing to minimize the negative log-likelihood of the ground-truth token sequence.
Given a dataset of paired audio–symbolic examples $\{(x^{(i)}, \mathbf{y}^{(i)})\}_{i=1}^{M}$, the training loss is defined as
\begin{equation}
\mathcal{L}(\theta) = - \sum_{i=1}^{M} \sum_{n=1}^{N^{(i)}}
\log p\big(y^{(i)}_n \mid y^{(i)}_{<n}, \mathbf{H}^{(i)}\big),
\end{equation}
where $\mathbf{H}^{(i)} = \mathrm{Enc}(x^{(i)})$ and $N^{(i)}$ is the length of the target token sequence. This corresponds to a token-level cross-entropy loss over the vocabulary $\mathcal{V}$.
\section{Experimental Evaluation}
\label{sec: experimental_evaluation}
In this section, we report the results of the experimental evaluation of our ADT model. We first describe the training procedure, with particular emphasis on the design of the proposed synthetic data generation pipeline. We then present results to assess the effectiveness of our approach and to evaluate the contribution of each component of the proposed framework.
\subsection{Data Synthetic Generation}
\label{ssec: synthetic_generation}
The training data for our model is derived from the Lakh MIDI Dataset \cite{Raffel2016-vr}, using the LMD-matched partition, a collection of 45,129 files aligned with the Million Song Dataset to ensure high-quality metadata.  Our model is trained exclusively on synthetic audio generated on-the-fly from the LMD MIDI files. For the syntetization strategy, we follow the same procedure of \cite{Sato2023-ry}. For each training step, a 2.56-second segment is randomly selected from a MIDI file and rendered into audio by a synthesizer that utilizes our semi-supervised one-shot library. Following \cite{Sato2023-ry} to augment the dataset and increase the diversity of the training material, we generate new synthetic samples by linearly interpolating between pairs of existing sounds. Let \(x_1^{(c_i)}\) and \(x_2^{(c_i)}\) be two distinct audio samples belonging to the same instrument category \(c_i\). We define a new interpolated sample \(x_{\text{new}}^{(c_i)}\) as:
\[
x_{\text{new}}^{(c_i)} = \alpha x_1^{(c_i)} + (1 - \alpha) x_2^{(c_i)}
\]
where the mixing coefficient \(\alpha\) is drawn from a uniform distribution, \(\alpha \sim \mathcal{U}(0, 1)\). 
\subsection{Results}
\label{ssec:subhead}
\subsubsection{Evaluation data}
\label{sssec: test_dataset}
Following previous work in ADT \cite{Vogl2018-wj, Cartwright2018-rj, Zehren2021-ht}, we evaluate our model on the MDB \cite{inproceedings} and ENST \cite{Gillet2006-jv} drum datasets. These datasets contain professionally recorded performances by drummers playing across a variety of styles and tempi, and include both drum-only tracks and recordings with melodic accompaniment. In this study, however, we restrict our evaluation to the DTD setting, i.e., drum-only recordings, and compare our results with prior work under this same evaluation protocol.
Although our system is capable of disambiguating up to 26 instrument classes, for the sake of comparability with previous work, we conduct the evaluation by mapping our 26 labels to the 8-instrument taxonomy proposed in \cite{Zehren2021-ht}. Since their labeling scheme is also derived from MIDI note numbers defined in the General MIDI percussion key map, we directly adopt their mapping to obtain a reduced set of 8 instrument classes from our original 26-class vocabulary.
\subsubsection{Experimental settings}
\label{sssec: exp_settings}
\begin{table}[htbp]
\centering
\caption{F1-score comparison across experimental settings.
Best results per category are in \textbf{bold}.}
\label{tab:results_settings}
\resizebox{\columnwidth}{!}{%
\begin{tabular}{ll|cccccc}
\toprule
 &  & \multicolumn{6}{c}{F1-SCORE} \\
Test Dataset & Setting
& SUM & BD & SD & TT & HH & CY+RD \\
\midrule
\multirow{4}{*}{ENST}
 & Setting-1
 & \textbf{0.73} & \textbf{0.87} & \textbf{0.80} & 0.55 & \textbf{0.77} & 0.49 \\
 & Setting-2
 & 0.67 & 0.84 & 0.77 & 0.43 & 0.75 & 0.38 \\
 & Setting-3
 & 0.70 & 0.85 & 0.75 & 0.42 & 0.73 & \textbf{0.52} \\
 & Setting-4
 & 0.69 & 0.83 & 0.75 & \textbf{0.58} & 0.69 & 0.50 \\
\midrule
\multirow{4}{*}{MDB}
 & Setting-1
 & \textbf{0.79} & 0.92 & \textbf{0.85} & \textbf{0.77} & 0.74 & \textbf{0.52} \\
 & Setting-2
 & 0.73 & 0.91 & 0.79 & 0.45 & 0.72 & 0.43 \\
 & Setting-3
 & 0.76 & \textbf{0.93} & 0.83 & 0.46 & 0.73 & 0.44 \\
 & Setting-4
 & \textbf{0.79} & 0.91 & \textbf{0.85} & 0.75 & \textbf{0.77} & \textbf{0.52} \\
\bottomrule
\end{tabular}%
}
\end{table}

As mentioned earlier, we conduct a systematic evaluation to assess the effectiveness of our approach. We define four experimental settings in which the ADT system is trained under different conditions to validate the main components of the proposed framework independently. The results of each setting is reported in Table~\ref{tab:results_settings}.
We begin by defining a baseline configuration (Setting~1), in which the model is trained using the CLAP-augmented one-shot library and the full vocabulary of 26 instrument classes, including tempo and velocity information in the token sequence. 
\begin{figure*}[t]
    \centering
    \includegraphics[
        width=\textwidth,
        height=0.485\textheight,
        keepaspectratio
    ]{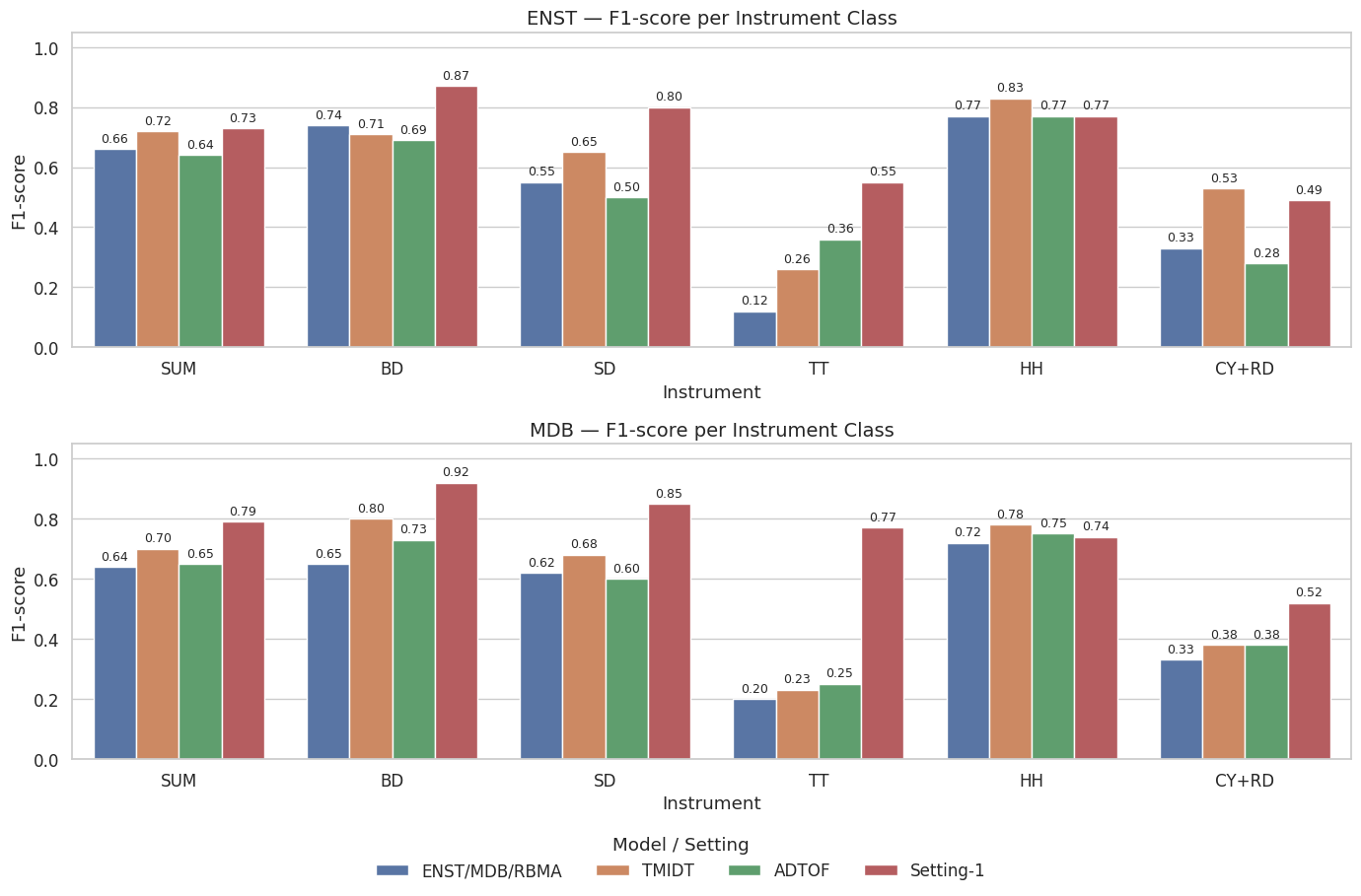}
    \caption{Metrics reported for the 3-split cross-validation on ENST and MDB dataset.}
    \label{fig:metrics_chart}
\end{figure*}
Figure~\ref{fig:metrics_chart} compares the performance of our baseline with the results reported in \cite{Vogl2018-wj} and \cite{Zehren2021-ht}. Our sequence-to-sequence model trained on a synthetic distribution outperforms all previously proposed systems.
We then evaluate the impact of the proposed data construction strategy by training the model using only the manually curated, supervised gold set of one-shot samples, while keeping all other hyperparameters fixed. The results of this experiment are reported in Setting~2 and provide clear evidence that the model benefits from the larger dataset constructed through our semi-supervised pipeline.
To further validate our approach, we conduct an additional experiment in which we retain the same set of one-shot samples obtained via the CLAP matching strategy, but reassign instrument labels using a naming-based heuristic. Specifically, we employ a normalized Levenshtein distance to match instrument labels based on the filename and parent directory name of each one-shot sample. This procedure results in an alternative version of the one-shot library that differs only in the labeling strategy. The results of this experiment, reported in Setting~3, show that the CLAP-based labeling consistently outperforms the naming-based approach (see Table~\ref{tab:results_settings}), providing better label coherence and improved class disambiguation for the transcription model.
Finally, we investigate the effect of the proposed 26-class instrument vocabulary. Following the same label mapping adopted for evaluation, we reduce the number of instrument classes in the one-shot library to eight and train a model using this reduced vocabulary. The results of this experiment are reported in Setting~4 and indicate that the model benefits from the finer-grained instrument categorization used in the baseline, supporting the adoption of the proposed 26-class vocabulary as a suitable standard for the ADT task.
\section{Conclusion}
\label{sec:conclusion}
In this paper, we introduced a novel data-centric paradigm for Automatic Drum Transcription that eliminates the need for paired audio–MIDI training data. By leveraging a semi-supervised pipeline based on CLAP embeddings, we constructed a large, diverse, and standardized corpus of one-shot drum samples from unstructured audio sources. This corpus enabled the synthesis of high-quality training data from MIDI files alone, effectively reducing the synthetic-to-real domain gap commonly observed in prior approaches. Combined with a sequence-to-sequence Transformer architecture and a principled tokenization strategy, our method achieves new state-of-the-art results on the ENST and MDB benchmarks, outperforming both fully supervised systems and previous synthetic-data-based methods.
\section{Acknowledgments}
\label{sec:Acknowledgments}
This work has been carried out while Pierfrancesco Melucci
was enrolled in the Italian National Doctorate on Artificial Intelligence run by Sapienza University of Rome in collaboration
with Universita di Roma Tre.

\bibliography{references}
\end{document}